\documentclass[12pt,preprint]{aastex}

\makeatother

\lefthead{Magdis et al..~}

\slugcomment{to appear in ApJ}
\shorttitle{Herschel/PACS observations of z$\sim$3 LBGs}
\shortauthors{Magdis et al.}
\begin{document}
 \title{A First Glimpse into the far-IR properties of high-z UV-selected Galaxies: Herschel/PACS observations of z$\sim$3 LBGs}
 \author{G.E Magdis $\!$\altaffilmark{1},
 	     D.Elbaz $\!$\altaffilmark{1},
         H.S. Hwang $\!$\altaffilmark{1},
         E. Daddi $\!$\altaffilmark{1},
         D. Rigopoulou $\!$\altaffilmark{2},
         B. Altieri $\!$\altaffilmark{3},
         P. Andreani $\!$\altaffilmark{4},
         H. Aussel $\!$\altaffilmark{1},
	     S. Berta $\!$\altaffilmark{5},
         A. Cava $\!$\altaffilmark{6},
         A. Bongiovanni $\!$\altaffilmark{6},
  		 J. Cepa $\!$\altaffilmark{6},
         A. Cimatti $\!$\altaffilmark{7},
         M. Dickinson $\!$\altaffilmark{8},
         H. Dominguez $\!$\altaffilmark{9},
         N. F{\"o}rster Schreiber $\!$\altaffilmark{5},
         R. Genzel $\!$\altaffilmark{5},
         J-S Huang 	$\!$\altaffilmark{15},
         D. Lutz $\!$\altaffilmark{5},
         R. Maiolino $\!$\altaffilmark{12},
		 B. Magnelli $\!$\altaffilmark{5},
		 G.E. Morrison $\!$\altaffilmark{10,11},
	     R. Nordon $\!$\altaffilmark{5},
	     A.M. P{\'e}rez Garc{\'i}­a $\!$\altaffilmark{6},
		 A. Poglitsch $\!$\altaffilmark{5},
		 P. Popesso $\!$\altaffilmark{5},
         F. Pozzi $\!$\altaffilmark{12},
         L. Riguccini $\!$\altaffilmark{1},
         G. Rodighiero $\!$\altaffilmark{13},
A. Saintonge $\!$\altaffilmark{5},
P. Santini $\!$\altaffilmark{12},
M. Sanchez-Portal $\!$\altaffilmark{5},
L. Shao $\!$\altaffilmark{5},
E. Sturm $\!$\altaffilmark{5},
L. Tacconi $\!$\altaffilmark{5},
I. Valtchanov $\!$\altaffilmark{14}
}
\altaffiltext{1}{CEA, Laboratoire AIM, Irfu/SAp, F-91191 Gif-sur-Yvette, France}
\altaffiltext{10}{Institute for Astronomy, University of Hawaii, Honolulu, HI 968226}
\altaffiltext{11}{Canada-France-Hawaii Telescope, Kamuela, HI 96743}
\altaffiltext{8}{NOAO, 950 N. Cherry Avenue, Tucson, AZ 85719, USA}
\altaffiltext{2}{Department of Astrophysics, Oxford University, Keble Road, Oxford, OX1 3RH}
\altaffiltext{3}{European Space Astronomy Centre, Villafrance del Castillo, Spain}
\altaffiltext{4}{ESO, Karl-Schwarzschild-Str. 2, D-85748 Garching, Germany}
\altaffiltext{5}{Max-Planck-Institut f\"{u}r Extraterrestrische Physik (MPE), Postfach 1312, 85741 Garching, Germany}
\altaffiltext{6}{Instituto de Astrof{\'i}sica de Canarias, 38205 La Laguna, Spain}
\altaffiltext{7}{Dipartimento di Astronomia, Universit{\`a} di Padova, Vicolo dell'Osservatorio 3,Italy}
\altaffiltext{9}{INAF-Osservatorio Astronomico di Bologna, via Ranzani 1, I-40127 Bologna, Italy}
\altaffiltext{12}{INAF - Osservatorio Astronomico di Roma, via di Frascati 33, 00040 Monte Porzio Catone, Italy}
\altaffiltext{13}{Dipartimento di Astronomia, Universit{\`a} di Padova, Vicolo dell'Osservatorio 3,35122 Padova, Italy}
\altaffiltext{14}{Herschel Science Centre}
\altaffiltext{15}{Harvard-Smithsonian Center for Astrophysics, 60 Garden Street, Cambridge, MA 02138, USA}
\begin{abstract}
We present first insights into the far-IR properties for a sample of
    IRAC and MIPS-24$\mu$m detected Lyman Break Galaxies (LBGs) at $z$ $\sim$ 3, as
    derived from observations in the northern field of the Great
    Observatories Origins Survey (GOODS-N) carried out with the PACS
    instrument on board the Herschel Space Observatory. Although none of
    our galaxies are detected by Herschel, we employ a stacking technique 
    to construct, for the first time, the average spectral energy distribution of  
    infrared luminous LBGs from UV to radio wavelengths. We derive a median IR luminosity of
    $L_{\rm IR}$ = 1.6 $\times$ 10$^{12}$ $L_\odot$, placing the population
    in the class of ultra luminous infrared galaxies
    (ULIRGs). Complementing our study with existing multi-wavelength
    data, we put constraints on the dust temperature of the population
    and find that for their $L_{\rm IR}$, MIPS-LBGs are warmer than
    submm-luminous galaxies while they fall in the locus of the $L_{\rm
      IR}$-$T\rm _{d}$ relation of the local ULIRGs. This, along with
    estimates based on the average SED, explains the marginal detection
     of LBGs in current sub-mm surveys and suggests that these
    latter studies introduce a bias towards the detection of colder
    ULIRGs in the high-$z$ universe, while missing  high-$z$ ULIRGS with warmer dust.
\end{abstract}

\section{Introduction}
Lyman break galaxies (LBGs), represent a significant fraction of z $\sim$ 3  
star-forming galaxies that have been discovered up to now. Since LBGs are 
the most common population at this epoch, investigating their properties 
is crucial to enhance our view of the early universe. A large number of multi-wavelength studies, spanning from X-rays to radio, have characterized the nature of the population. One recent advance into our understanding of this population was provided by the Infrared Array Camera (IRAC), the Multi-band Imaging Photometer (MIPS) and the Infrared Spectrograph (IRS) on board the Spitzer Space Telescope (Spitzer). In particular, IRAC measurements at rest-frame near-infrared wavelengths (3.6-8$\mu$m) indicate that their stellar masses are typically 10$^{9}$-10$^{11}$M$_\odot$ (e.g., Magdis et al. 2010a) while the detection of a fraction ($\sim$20\%) of these UV-selected galaxies at MIPS-24$\mu$m unveiled the sub-population of the infrared luminous Lyman break galaxies (Huang et al. 2005, Rigopoulou et al. 2006). Subsequent follow-up IRS spectroscopy of MIPS detected LBGs, revealed prominent Polycyclic Aromatic Hydrocarbon (PAH) features in their mid-IR spectrum, providing evidence for the existence of significant amounts of dust and that they are mainly powered by star-formation rather than an AGN (Huang et al. 2007, Rigopoulou et al. 2010). 

The star-formation rate (SFR) of MIPS detected LBGs has been studied in detail by
Magdis et al. (2010b), where they found that UV, mid-IR and radio measurements of the SFR for UV-selected galaxies are in good agreement, indicating an average $<$SFR$>$=250$M_{\odot} yr^{-1}$. This SFR along with the fact that these galaxies contain considerable amounts of dust as indicated by their mid-IR emission, suggests that MIPS-LBGs should be detectable at sub-mm wavelengths. Although the first attempts to detect far-IR counterparts were not successful (e.g. Chapman et al. 2000), recently there has been a significant progress in this front. Chapman et al. (2009) reported the SCUBA-850$\mu$m sub-mm detection of Westphal-MM8 while Rigopoulou et al. (2010) reported MAMBO 1.2mm detections of a further two LBGs, EGS-D49 and EGS-M28 selected based on their strong MIPS 24$\mu$m emission. In addition to those, two more lensed-LBGs 
have been detected in the sub-mm by Baker et al. (2001) (MS1512-cB58) and by Siana et al. (2009) (Cosmic Eye). Despite these promising detections though, the FIR properties of LBGs still remain elusive.

In this paper we use observations of the northern field of the Great Observatories Origins Survey (GOODS-N) obtained by Photoconductor Array Camera and Spectrometer (PACS)  (Poglitsch et al. 2010) on board the Herschel Space Observatory (Pilbratt et al. 2010). We complement our study with multi-wavelength data and we aim to get a first insight into the FIR properties of a sample of MIPS 24$\mu$m detected LBGs, construct the average spectral energy distribution (SED) of the population, derive estimates of their dust temperature ($T\rm _{d}$) 
and compare their properties to those of SMGs. The Herschel data used in this study are part of the Herschel Science Demonstration Phase of the PACS Evolutionary Probe (PEP, PI Lutz) program. Throughout this paper we assume $\Omega_{\rm m}$=0.3, H$_{0}$=71km sec$^{-1}$ Mpc$^{-1}$ and 
$\Omega_{\Lambda}$=0.7


\section{Sample Selection and Herschel Observations}
GOODS-North was observed by PACS, at 100- and 160$\mu$m in the framework of
the Guaranteed Time Key Program, Pacs Evolutionary Probe (PEP). The area covered by PACS was 10'$\times$15' in a total exposure time of 30 hours and an average integration time of $\sim$ 2.5sec per pixel. For source extraction two independent methods were employed: the traditional blind source extraction that was performed by using the Starfinder Point-spread function (PSF)-fitting code (Diolaiti et al. 2000) and a guided extraction using 24$\mu$m priors. In both methods, PSFs were extracted from the final science maps and were subsequently used for a PSF-fitting photometric analysis. Flux uncertainties were obtained by Monte-Carlo simulations, adding fake sources in the real maps. The measured 1$\sigma$ noise is 1.00 mJy at 100$\mu$m and 1.90 mJy at 160$\mu$m. For more details we refer the reader to the appendix of Berta et al. (2010). 

Our sample of LBGs is identical to the one studied by Magdis et al. (2010b) and we refer the reader to this paper for a detailed description. In brief we consider 49 IRAC detected ([3.6]$\rm_{AB}<$25.0 or 0.36$\mu$Jy) LBGs in GOODS-N, that have originally been optically selected (U,G,R, R$<$25.5) by Steidel et al. (2003). All galaxies in our sample have spectroscopic redshift (median z=2.95) derived by optical spectroscopy that has also been used to determine the absence of AGN signatures (i.e., strong high ionization emission lines) in their rest-frame UV spectrum (Steidel 2003, Shapley 2003). Among the original sample, 9 LBGs are also detected at MIPS-24$\mu$m (5$\sigma\label{•} \sim$20$\mu$Jy) with a median 24$\mu$m flux density $f_{\rm 24}$=31.2$\mu$Jy. Henceforth, we will refer to these LBGs as the MIPS-LBGs while to the remaining (40) as IRAC-LBGs.
IRAC photometry for these objects is provided in Magdis et al. (2010a), while an analysis of MIPS 24$\mu$m data is described Magdis et al. (2010b).
Matching our sample with the 100- and 160$\mu$m blind and prior catalogs returned no individual detections at the 3 sigma level (3mJy and $\sim$5mJy respectively). Since none of our LBGs is individually detected we use the residual maps to do stacking. On stacking we first considered the sample of MIPS-LBGs. We employed median stacking analysis, cutting sub-images of the residual maps centered at the optical position of each undetected LBG. The residual maps were created by Starfinder, by subtracting all individually detected sources down to 3$\sigma$. To avoid contaminating the stacked signal from residuals, we only added galaxies to the stack if there were no bright sources within $\sim$4''(100$\mu$m) and $\sim$6''(160$\mu$m). Then a stacked flux was measured in a manner similar to the measurement of the detected PACS sources. To quantify the error of our measurement we stacked at 9 random positions and repeated it 50.000 times. The 1$\sigma$ of the distribution of the derived fluxes was adopted as the uncertainty of our measurement. Stacking at 100$\mu$m returned no detection (f$_{100}$=0.47$\pm$0.36 mJy) indicating a 3$\sigma$ upper of f$_{100}<$1.1mJy, while at 160$\mu$m we recovered a median flux density $f_{160}$=2.21 $\pm$ 0.52mJy (S/N$\sim$4.2) for the MIPS-LBGs. The final 160$\mu$m stacked image is shown in Figure 1 along with the distribution of the fluxes that correspond to random stacked positions. This distribution indicates the probability to recover a stacked flux density $f_{\rm 160}>$ 4.2mJy by chance, is 2.31$\times$10$^{-5}$. Stacking the whole sample of LBGs, or just the IRAC LBGs alone, returned no detection in either PACS band. For the IRAC LBGs we determine a 3$\sigma$ upper limit of f$_{100}<0.47$mJy and f$_{160}<0.79$mJy while the actual fluxes that correspond to the stacked images are $f_{100}$=0.21$\pm$0.16 mJy and  f$_{160}$=0.32$\pm$ 0.26 mJy.  

For the IRAC sample we also consider median stacked flux densities of f$_{\rm 1.1mm}$=0.41$\pm$0.11mJy and f$_{\rm 1.4GHz}$=3.6$\pm$0.8$\mu$Jy as derived from stacking at 1.1mm AzTEC and 1.4GHz VLA map, while for the MIPS sample we consider a median stacked flux 
density of f$_{\rm 1.4GHz}$=8.5$\pm$2.2$\mu$Jy and a 3$\sigma$ upper limit of f$_{\rm 1.1mm}$=1.1mJy (Magdis et al. 2010b). We also add optical (ground based UGR, ACS BViz) and near-IR (J,K) photometry provided by the GOODS team.

\section{Results and Discussion}
\subsection{Spectral Energy Distribution of LBGs}
Using the multi-wavelength photometry (and the upper limits) described above we attempt to construct the full, (rest-frame UV to radio) average SED of MIPS selected LBGs. We fit the optical to near-IR part with model SEDs generated by the Charlot and Bruzual 2007 (CB07) code and the mid-IR to radio with template SEDs from the Chary \& Elbaz 2001 (CE01) and Dale \& Helou 2002 (DH02) libraries. For the fitting procedure we allow for renormalization of the CE01 templates while for the DH02 models the luminosities are normalized as described by Marcillac et al. (2006). Results based on the two methods are in very close agreement indicating a median L$_{\rm IR}$=1.6($\pm$0.5)$\times$10$^{12}$L$_{\odot}$ and that MIPS-LBGs belong to the class of Ultra-Luminous Infrared Galaxies (ULIRGs). This luminosity translates to SFR$=275M_{\odot}yr^{-1}$ (Salpeter IMF), in good agreement with the UV (250$^{+35}_{-80}$M$_\odot$yr$^{-1}$) and radio (280$\pm$85M$_\odot$yr$^{-1}$) SFR estimates presented in Magdis et al. (2010b). Adopting a median stellar mass of MIPS LBGs M$_{\ast}$=7.9$\times10^{10}\rm M_{\odot}$ (Magdis et al. 2010a), we derive a specific star-formation rate of SSFR=3.5Gyr$^{-1}$. This value is lower than that derived 
for the IRAC-LBGs (4.3Gyr$^{-1}$, Magdis et al. 2010a) but higher than that found for z$\sim$2 star-forming galaxies (2.5 Gyr$^{-1}$, Daddi et al. 2007a). Finally, based on the best fit CE01 model, we estimate a median $f_{\rm 850}$=1.36mJy, very close to the confusion limit of current sub-mm surveys. We comment on this finding later on where we present a comparison between LBGs and SMGs.     

\subsection{Indication of Warmer Dust in MIPS-LBGs}
As we have already discussed, stacking IRAC-LBGs at 1.1mm has recovered a flux density of $f_{1.1mm}$=0.41$\pm$0.1mJy while at 160$\mu$m there is no detection. The situation is exactly the opposite for the MIPS-LBGs where stacking at 1.1mm indicate no detection while at 160$\mu$m we recover a median flux of $f_{160}$=2.21$\pm$0.52 mJy. This inversion could be interpreted as a change in the shape of the SED of the two samples. Furthermore a detailed study of IRAC-LBGs by Magdis et al. (2010b), indicated that the average IRAC detected LBG is a Luminous Infrared Galaxy (LIRG) with $L_{\rm IR}$=4.5($\pm$1.2)$\times$10$^{11}$ $L_{\odot}$, less luminous than the current sample of MIPS-LBGs. It has been observed that in the local universe ULIRGs have warmer dust and their SEDs peak at shorter wavelengths than those of LIRGs. Here we attempt to put constraints on the peak of the SED and the dust temperature ($T_{\rm d}$) of LBGs and 
investigate these parameters as a function of luminosity at z$\sim$3. 

Although ideally such a task requires photometric points close to the peak of the SED of a galaxy 
(which in this case translates into SPIRE data), here we use PACS 160$\mu$m and 
AzTEC 1.1mm flux densities and upper limits, with the advantage of sampling 
both sides around the peak of the SED. Using 3$\sigma$ upper limits for the $f_{160}$ of IRAC-LBGs could 
provide us with an upper limit of the $T_{\rm d}$ of the sample while 3$\sigma$ upper limits for the $f_{\rm 1.1mm}$ of the MIPS LBGs will yield an estimate of a lower limit of the $T_{\rm d}$ of the sample.
 
We first employed Monte Carlo simulations to generate 10.000 combinations of 160$\mu$m and 1.1mm fluxes. 
For each of the two bands, the simulated fluxes were generated following the flux distribution derived from the stacking simulations and centered on the measured stacked flux. 
To derive the dust temperature for each of the generated realizations, we used a single temperature greybody fitting form, $F_{\nu} \propto \nu^{3+\beta}/(exp(h\nu/kT_{d})-1)$, 
where $\beta$ is the dust emissivity and T$_{\rm d}$ is the effective or emission weighted dust temperature. We fixed the $\beta$ value to 1.5 (Gordon et al. 2010) and considered the median redshifts as representative for each sample (z=2.98 for IRAC-LBGs and z=2.92 for the MIPS-LBGs). The $T_{\rm d}$ for each realization was then obtained from the best fit model that was derived based on the minimization of the $\chi^{2}$ value. Since at 100$\mu$m (rest-frame 25$\mu$m) there is strong contribution of Very Small Grains (VSGs) in the IR emission, $f_{100}$ upper limits were not fitted but they were considered as a sanity check, i.e. that the models do not violate the upper limits. Based on the distribution of the obtained dust temperatures (Figure 3) we then derive an upper limit of $T_{\rm d}$ = 38.2(39.3) K (IRAC-LBGs), and a lower limit of $T_{\rm d}$ = 40.1(38.9) K (MIPS-LBGs) at a 2$\sigma$(3$\sigma$) confidence level. The best fit models that correspond to the 2$\sigma$ temperature limits are depicted in Figure 4. We note that adopting $\beta$ = 2.0 instead of 1.5, would result in a lower dust temperature, by 2-3 K.

The above analysis indicates that the dust temperatures of the two samples are different at a 2.6$\sigma$ confidence level, with MIPS-LBGs having warmer dust. Since MIPS-LBGs and IRAC-LBGs are objects divided by a threshold in their rest 6$\mu$m emission, which is due to warm, very small grains and/or PAHs, if the presence of warm small grains correlates with 
the presence of warm larger grains then our result would be object of selection bias. To explore this possibility, we consider a sample of local ULIRGs with ISOCAM 6.75$\mu$m and IRAS 60- and 100$\mu$m data (Elbaz et al. 2002). Plotting the $f_{\rm 100}/f_{\rm 60}$ which is a dust temperature indicator, against the rest-frame 6.75$\mu$m luminosity (that corresponds to 24$\mu$m emission of z$\sim$3), reveals that there is no obvious correlation between 6$\mu$m luminosity and $T_{\rm d}$. Hence, assuming that this applies for galaxies at z$\sim$3, it seems that our result is not subject to a selection bias. On the other hand, as we have already shown, MIPS-LBGs are more infrared-luminous than IRAC-LBGs with corresponding infrared luminosities L$_{\rm IR}$=1.6($\pm$0.5)$\times$10$^{12}$L$_{\odot}$ for MIPS- and L$_{\rm IR}$=4.5($\pm$0.5)$\times$10$^{11}$L$_{\odot}$ for IRAC-LBGs. Therefore, it appears that we observe for the first time the general $L_{\rm IR}$$-$$T_{\rm d}$ trend seen in the local universe, (with more luminous galaxies having warmer dust), for UV-selected galaxies at $z$ $\sim$ 3. 

\subsection{Comparison with SMGs}
Considering the large SFRs and substantial dust reddening that are inferred
for some LBGs, and the fact that their spectra exhibit characteristics of local starbursts (Pettini et al. 2001), it is somewhat surprising that there are only few examples of direct sub-millimeter detection for these galaxies. MIPS-LBGs are the most rapidly star-forming, most luminous, and dustiest
galaxies among the high redshift UV-selected population, and therefore are the best candidates
for having far-IR emission that could be detected in current sub-mm surveys. In what follows we discuss how the current data can shed light on the mystery of this marginal detection of LBGs in sub-mm bands. Based on the average SED of MIPS-LBGs that we presented here, 
we predict that the flux density of the MIPS-LBGs emitted at 850$\mu$m is $ f_{\rm 850}$=1.1-1.5mJy, just below the current confusion limit. 
It could therefore be suggested that MIPS-LBGs provide a link between SMGs and typical UV selected LBGs that are faint in the IR. It has also been speculated that the fact that LBGs are not detected in the sub-mm bands is due to warmer dust compared to that of SMGs (Kaviani et al. 2003). Although to test this one needs both robust measurements of the $T_{\rm d}$ (based on observations that probe the peak of the SED of the galaxies) and to have a large sample of LBGs, the upper limits that we derived in this study can provide a first insight.  

Figure 5 compares dust temperature versus infrared luminosity for the MIPS and IRAC LBGs as derived from this study. For comparison we also consider the large compilation of $z$ $\sim$ 2 SMGs by Chapman et al. 2005, and the local/intermediate-z samples of ULIRGs presented by Clements et al. (2010) and Yang et al. (2007). In all these studies, the $T_{\rm d}$ was measured in manner that is similar to ours, fitting modified black-body models with fixed $\beta$=1.5 to the FIR photometric points of each sample, hence the comparison between these studies and our sample is meaningful. We also plot, the $3\sigma$ envelope of the $L_{\rm IR}$-$T_{\rm d}$ relation for local infrared galaxies in SDSS (Hwang et al. 2010 in prep).

It is evident that for the $L_{\rm IR}$ of the MIPS-LBGs, the bulk of SMGs are considerably colder, 
while MIPS-LBGs fall in the locus of the local ULIRGs and are within the scatter observed in local galaxies. 
Based on modified black body models, we also compute tracks of constant 850$\mu$m flux density for galaxies at $z$=3, close to the confusion/detection limit of current sub-mm surveys ($f_{\rm 850}$=1mJy and $f_{\rm 850}$=2mJy). MIPS-LBGs lie in between the two tracks, indicating that a typical
MIPS detected LBG emits at 1-2 mJy level at the sub-mm bands, in excellent agreement with the median $f_{\rm 850}$ flux density derived from the average SED of the population. This explains the small overlap between the LBGs and SMGs found in previous studies.
For a given star formation rate and optical depth for dust absorption (and hence,
for a given bolometric far-infrared luminosity), the warmer dust temperature in LBGs
leads to fainter sub-mm emission compared to that found in typical SMGs.
This finding also suggests that star-formation in LBGs is spatially compact. On the other hand, recent studies based on minimal excitation (CO J=1 $\rightarrow$ 0) emission that traces the bulk of the metal-rich molecular gas provide evidence of a more extended distribution of star-formation in SMGs (e.g. Ivison et al. 2010), that could explain their colder dust component. Generalizing our results, we argue that ULIRGs with warmer dust than that of SMGs exist in the high-$z$ universe and would not be selected in current sub-mm surveys. This is further enforced by the study of IRAC selected ULIRGs at $z\sim$ 2, (Magdis et al. 2010 in prep), where based on Herschel PACS/SPIRE observations, we find 
that 60$\%$ of the sample would be missed by current sub-mm surveys.


We have presented a first glimpse into the far-IR properties of $z\sim$ 3 LBGs, demonstrating the power of Herschel in the study even of high-z UV-selected galaxies and its ability to address issues that we were unable to resolve in the pre-Herschel era. We aim to further extent this study, complementing PACS with SPIRE data that probe the peak of the SED of $z$ $\sim$ 3 galaxies and taking advantage of the Super-Deep and Ultra-Deep surveys in GOODS-N and GOODS-S respectively as part of the Herschel-GOODS program (PI D. Elbaz). 
 
\begin{acknowledgements}
PACS has been developed by a consortium of institutes led by MPE (Germany) and including UVIE (Austria); KUL, CSL, IMEC (Belgium); CEA, OAMP (France); MPIA (Germany); IFSI, OAP/AOT, OAA/CAISMI, LENS, SISSA (Italy); IAC (Spain). This development has been supported by the funding agencies BMVIT (Austria), ESA-PRODEX (Belgium), CEA/CNES (France), DLR (Germany), ASI (Italy), and CICT/MCT (Spain).
\end{acknowledgements}

\begin{figure*}
\centering
\includegraphics[width=0.35\textwidth,angle=0]{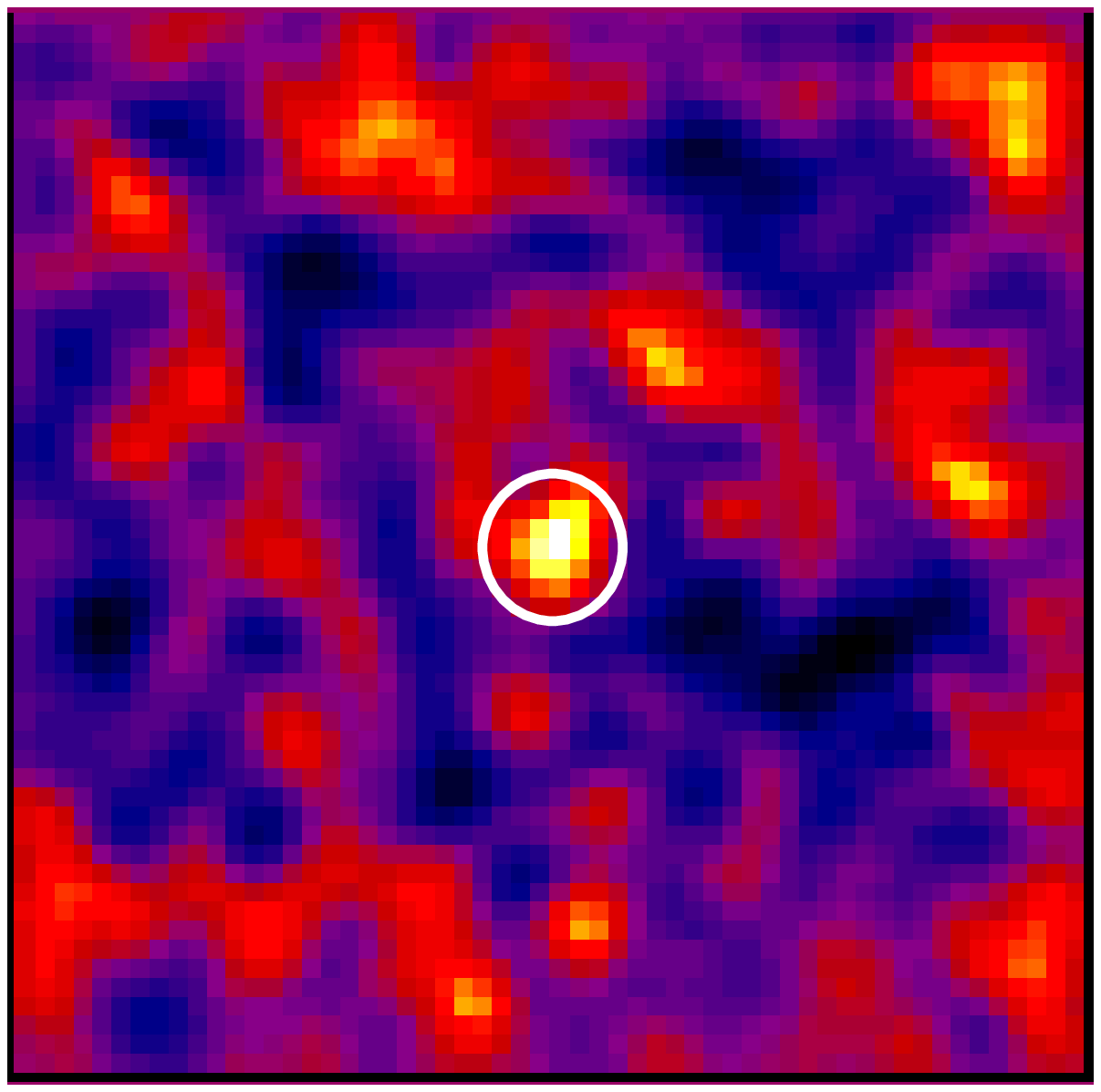}\hspace*{0.03\textwidth}
\includegraphics[width=0.35\textwidth,angle=0]{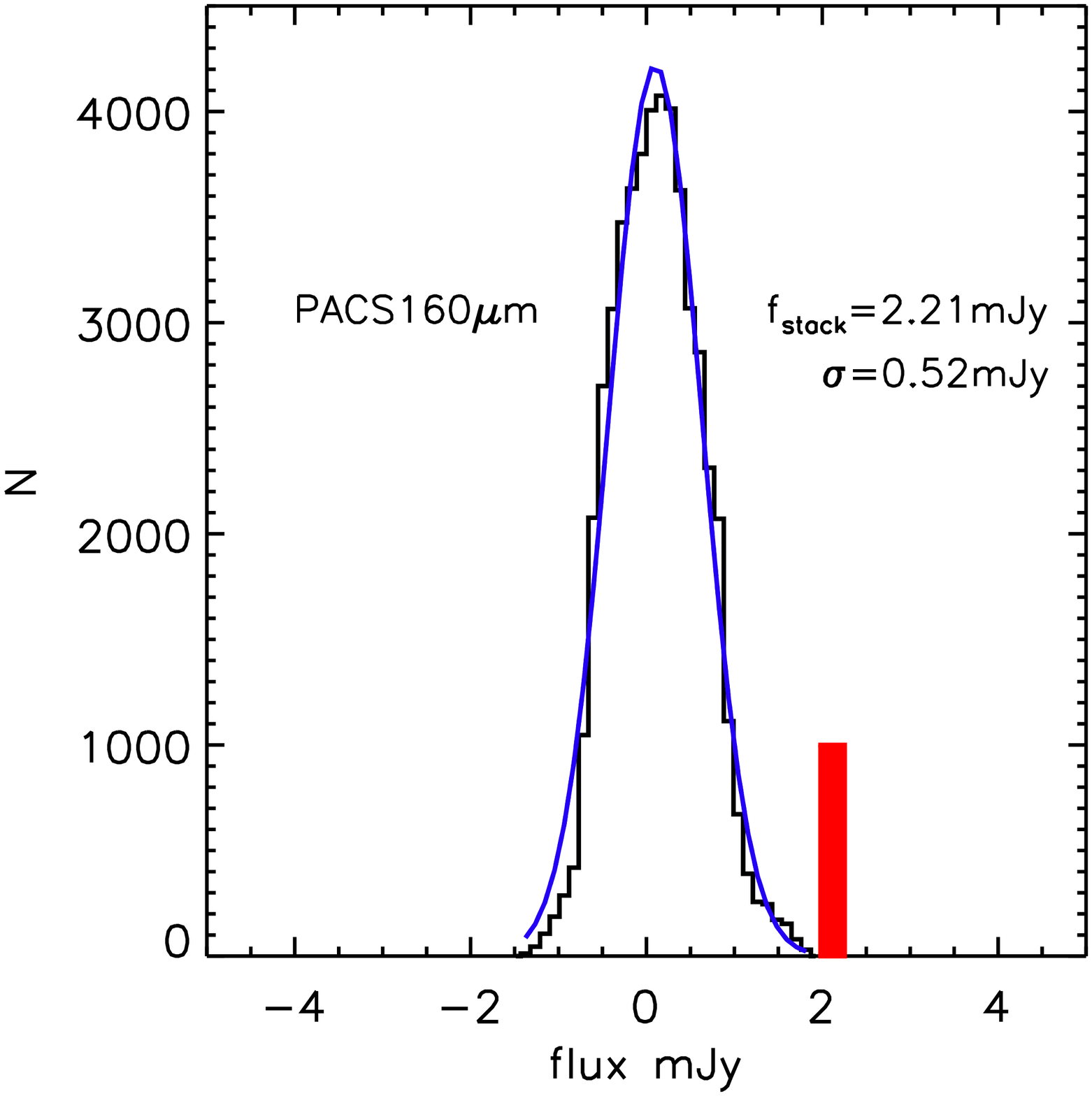}\hspace*{0.03\textwidth}
\caption{Left) PACS-160$\mu$m stacked image (130''$\times$130'') of 9 MIPS-24$\mu$m detected LBGs in GOODS-N. Right) Stacking simulations at 160$\mu$m. Distribution of the measured fluxes derived from 50000 stackings at 9 random positions along with the best Gaussian fit (rms=0.52mJy). The red line denotes the flux measurement of the stacking at the position MIPS-LBGs indicating a $\sim$4.2$\sigma$ detection.}

\label{fig:sub} %
\end{figure*}

   \begin{figure*}
   \centering
   \includegraphics[scale=0.6]{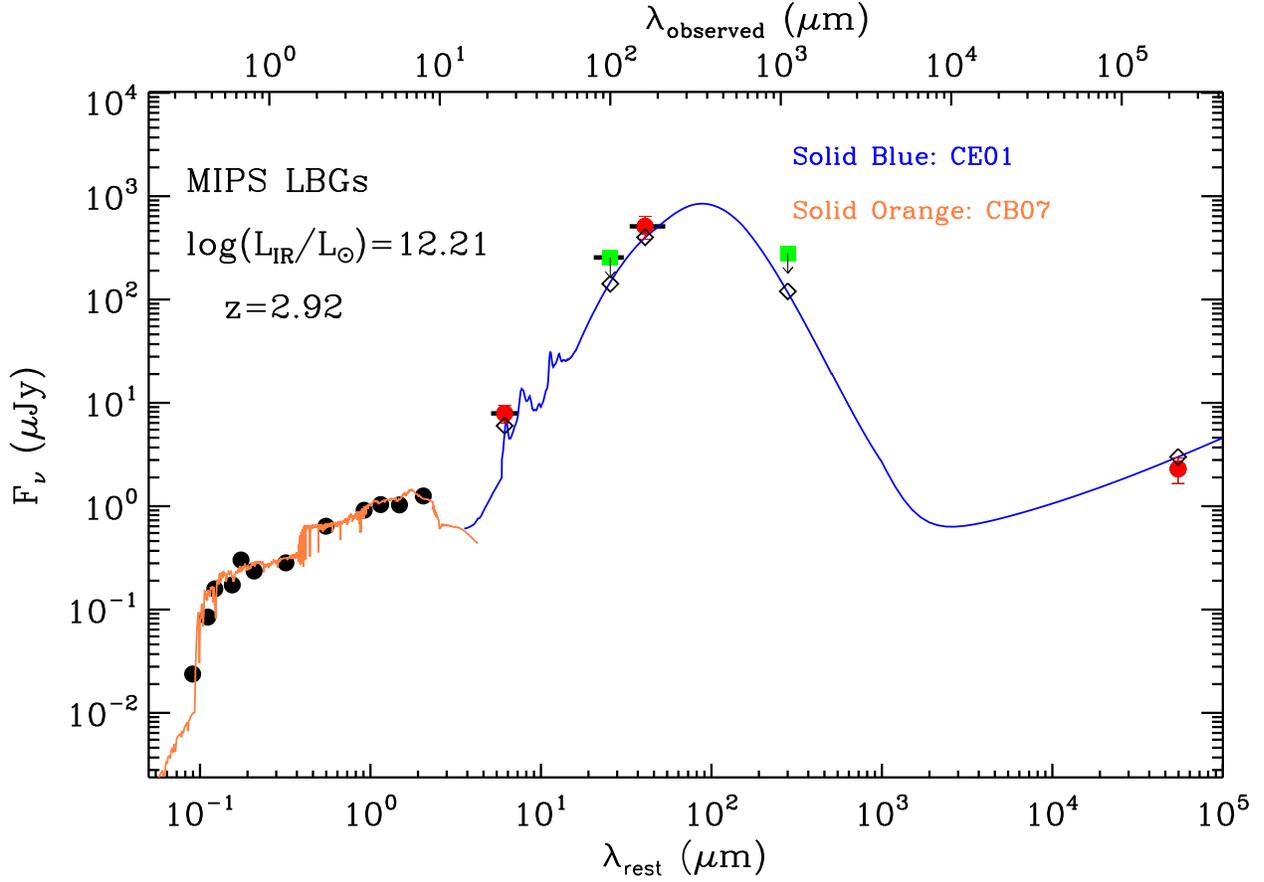}
   \caption{Rest--frame average SED of z$\sim$3, MIPS-LBGs. For the SED we use the median UGR+BViJK+IRAC+MIPS24 photometry of MIPS-LBGs, and the values (or upper limits) derived from stacking 100$\mu$m, 160$\mu$m AzTEC and radio. Upper limits are indicated with green squares. The rest-frame UV-NIR portion of the data is overlaid with the best fit CB07 model (orange line), while the mid-IR to radio is shown with the best--fit CE01 model (blue line). Black diamonds show the fluxes predicted from the best fit CE01 model.}
   
     \label{FigGam}%
    \end{figure*}       

\begin{figure*}
   \centering
   \includegraphics[scale=0.65]{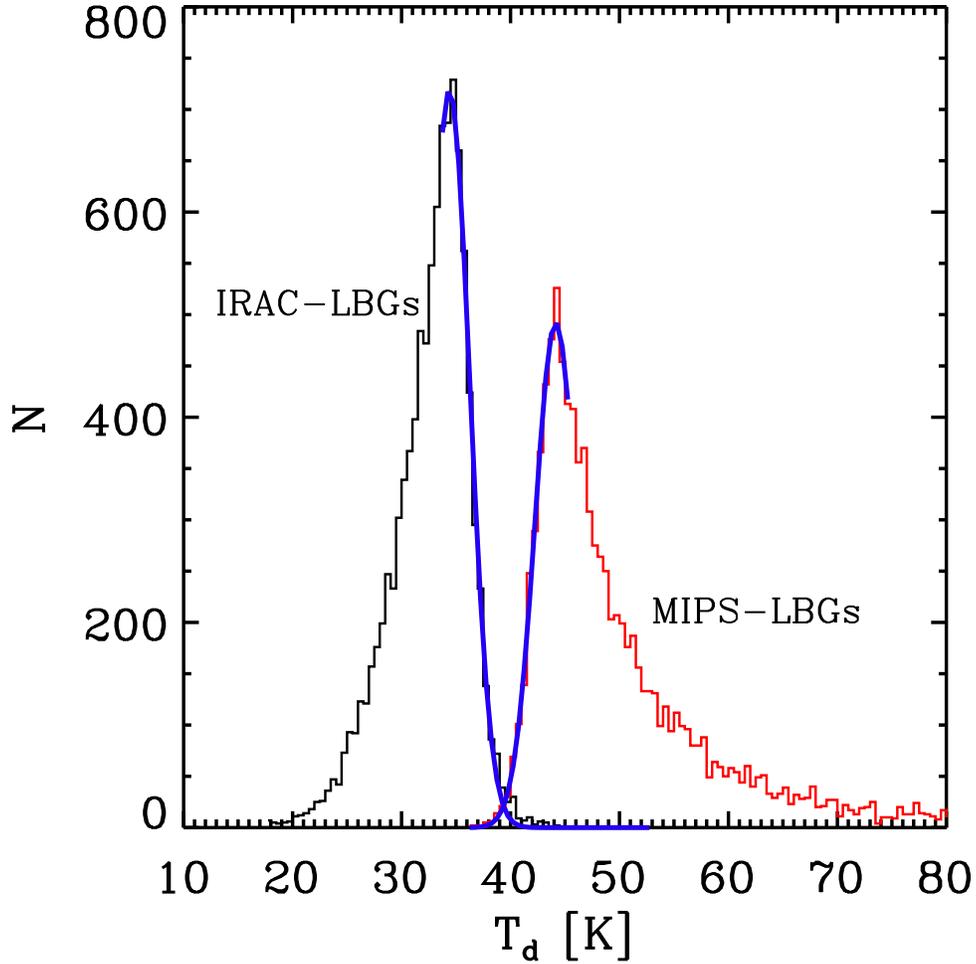}
   \caption{Distribution of the derived $T_{\rm d}$ for the IRAC- and MIPS-LBGs based on MC simulations. The two distributions begin to overlap at a 2.6$\sigma$ confidence level. Blue lines illustrate a gaussian fit to the left (MIPS-LBGs) and right (IRAC-LBGs) part of the distribution.}
    \label{FigGam}%
    \end{figure*}

\begin{figure*}
   \centering
   \includegraphics[scale=0.65]{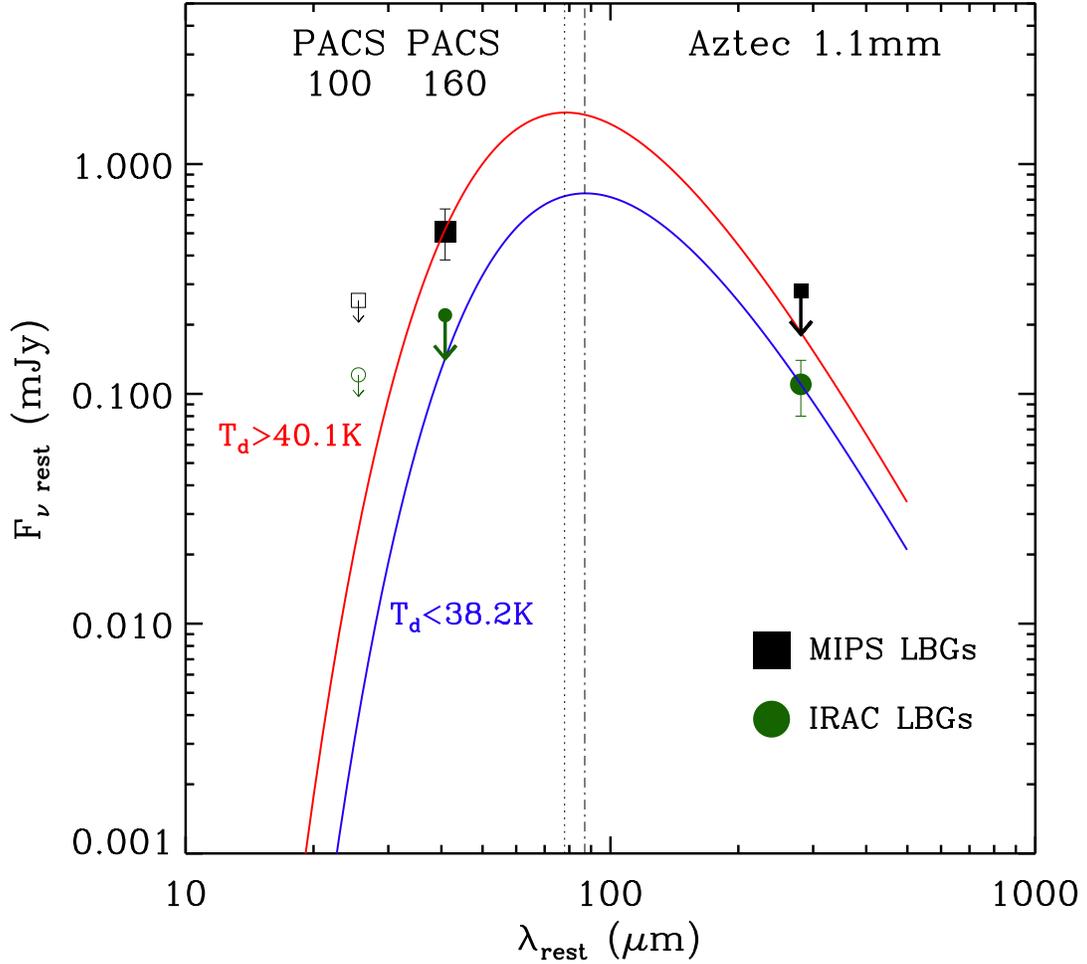}
   \caption{Derivation of lower and upper $T_{\rm d}$ limit estimates for MIPS- and IRAC-LBGs respectively. Filled black squares indicate the median stacked flux density of MIPS-LBGs at 160$\mu$m and the 3$\sigma$ upper limits of the flux density at 1.1mm as estimated from stacking. The solid red line shows the best fit modified black body model that corresponds to the 2$\sigma$ $T_{\rm d}$ lower limit. For the model we assume fixed $\beta$=1.5. Similarly, filled green circle indicate the median flux density of IRAC-LBGs at 1.1mm from Magdis et al. 2010b and the 3$\sigma$ flux upper limit at 160$\mu$m as derived from stacking in the present study. The solid blue line shows the best fit modified black body model that corresponds to the 2$\sigma$ T$_{\rm d}$ upper limit. The open black squares and green circle indicate 3$\sigma$ flux upper limit at 100$\mu$m as derived from stacking for MIPS- and IRAC-LBGs respectively. Vertical black dotted and dashed-dotted lines indicate the wavelength where the far-IR emission peaks for MIPS-LBGs and IRAC-LBGS.}
    \label{FigGam}%
    \end{figure*}

\begin{figure*}
   \centering
   \includegraphics[scale=0.65]{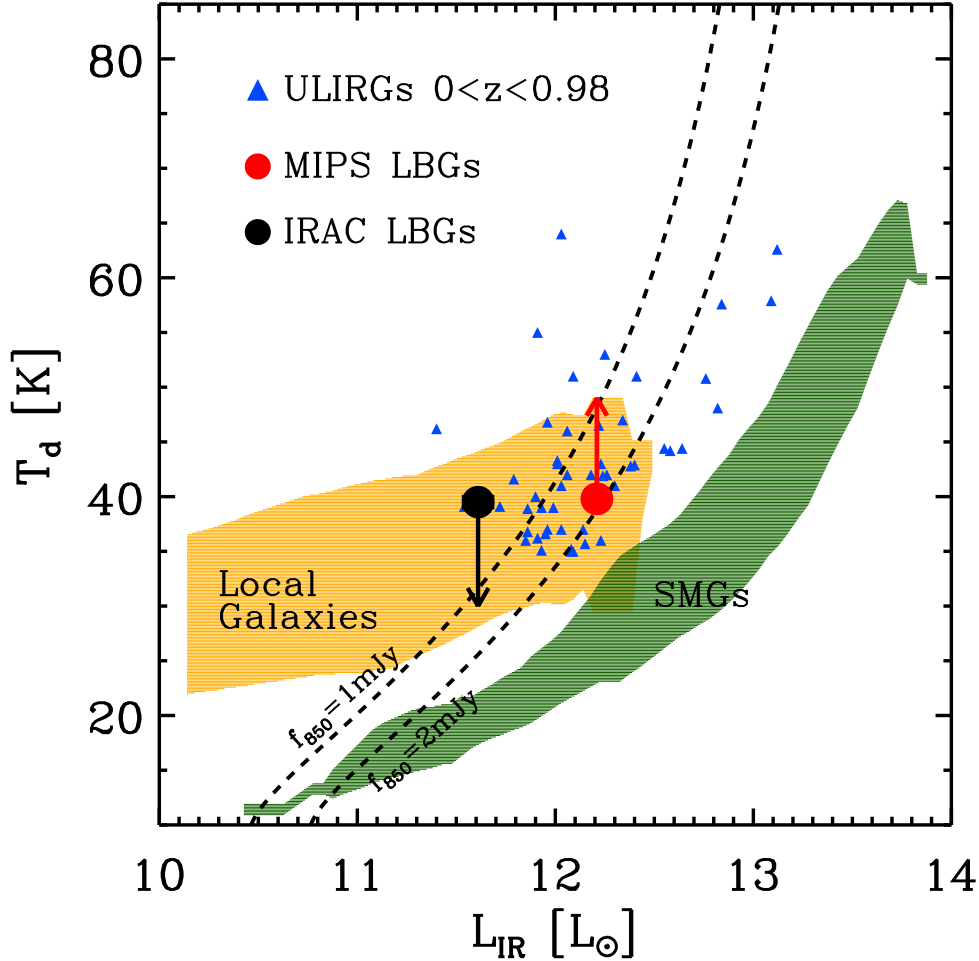}
   \caption{The $L_{\rm IR}-T_{\rm d}$ relation for MIPS and IRAC LBGs (red and black circles respectively). We note that these values correspond to 3$\sigma$ upper and lower $T_{\rm d}$ limits for IRAC and MIPS LBGs. Included are results for 0$<$z$<$0.98 ULIRGs (blue filled triangles, Clements et al. 2010, Yang et al. 2007). The green shaded region depicts the loci of high-z SMGs by Chapman et al. (2005), while the orange shaded area shows the 2$\sigma$ envelope of the $L_{\rm IR}-T_{\rm d}$ relation for local IR galaxies in SDSS (excluding AGNs) adopted from Hwang et al. (2010 in prep). Black dashed lines represent tracks of constant flux density at 850$\mu$m ($f_{\rm 850}$=1mJy and $f_{\rm 850}$=2 mJy) for galaxies at z=3. Objects at z=3 with higher $f_{\rm 850}$, lie on the right of the lines. This plot demonstrates that for a given $L_{\rm IR}$, the bulk of SMGs are considerably colder than MIPS-LBGs which fall in the locus of the local ULIRGs. We also note that for their $T_{\rm d}$ and $L_{\rm IR}$, MIPS-LBGs fall in between the 1- and 2 mJy constant flux density at 850$\mu$m, indicating that they would be missed by current ground-based sub-mm surveys. Finally, there is a hint that the LBGs at $z\sim$3 follow the general trend observed in the local universe, with more luminous galaxies having warmer dust.}
    \label{FigGam}%
    \end{figure*}

\end{document}